\documentclass[final,5p,times,twocolumn]{elsarticle}

\usepackage{graphicx}
\usepackage[amssymb,mediumqspace,thinspace]{SIunits}
\usepackage{xcolor}
\usepackage[normalem]{ulem}
\usepackage{amssymb,amsmath}
\usepackage{rotating}
\usepackage{adjustbox}
\usepackage{comment}

\usepackage{lineno}

\biboptions{sort&compress}
\modulolinenumbers[5]
\journal{arXiv (author accepted manuscript)}
\begin{document}

\begin{frontmatter}

\title{The role of dwell hold on the dislocation mechanisms of fatigue in a near alpha titanium alloy}

%% use the tnoteref command within \title for footnotes;
%% use the tnotetext command for the associated footnote;
%% use the fnref command within \author or \address for footnotes;
%% use the fntext command for the associated footnote;
%% use the corref command within \author for corresponding author footnotes;
%% use the cortext command for the associated footnote;
%% use the ead command for the email address,
%% and the form \ead[url] for the home page:
%%
%% \title{Title\tnoteref{label1}}
%% \tnotetext[label1]{}
%% \author{Name\corref{cor1}\fnref{label2}}
%% \ead{email address}
%% \ead[url]{home page}
%% \fntext[label2]{}
%% \cortext[cor1]{}
%% \address{Address\fnref{label3}}
%% \fntext[label3]{}

%% use optional labels to link authors explicitly to addresses:
%% \author[label1,label2]{<author name>}
%% \address[label1]{<address>}
%% \address[label2]{<address>}

\author[IC]{S. Joseph\corref{mycorrespondingauthor}}
\author[NH]{K. Joseph}
\author[IC]{T.C. Lindley}
\author[IC]{D. Dye\corref{mycorrespondingauthor}}
\cortext[mycorrespondingauthor]{mycorrespondingauthor}
\ead{david.dye@imperial.ac.uk}
\ead{sudha.joseph@imperial.ac.uk}

%\author{D.Dye\corref{firstcorr}\fnref{firstfoot}}
%\cortext[firstcorr]{david.dye@imperial.ac.uk}

%\author{S.Joseph\corref{secondcorr}\fnref{secondfoot}}
%\cortext[secondcorr]{sudha.joseph@imperial.ac.uk}

\address[IC]{Department of Materials, Royal School of Mines, Imperial College, Prince Consort Road, South Kensington, London SW7 2BP, UK}
\address[NH]{Department of Mathematics, New Horizon College of Engineering, Bangalore 560103, India}

\begin{abstract}
The dislocation structures appearing in highly mis-oriented soft/hard grain pairs in near-alpha titanium alloy Ti6242Si  were examined with and without the application of load holds (dwell) during fatigue. Dislocation pile-up in a soft grain resulted in internal stresses in an adjacent hard grain which could be relaxed by dislocation multiplication at localized Frank-Read sources, a process assisted by the provision of a relaxation time during a load hold.  The rate of this process is suggested to be controlled by $\langle c+a \rangle$ pyramidal cross-slip and $\langle a \rangle$ basal junction formation. A high density of $\langle a \rangle$ prism pile-ups was observed with dual slip on two prism planes, together with edge dislocations on the third prism plane in the soft grain of a highly mis-oriented grain pair, increasing the pile-up stress. The stress concentration developed by such pile-ups is found to be higher in dwell fatigue (single-ended pile-ups) than in LCF (double ended). Analytical modelling shows that the maximum normal stress produced on the hard grain in dwell fatigue by this pile-up would be near-basal, $\approx 2.5\degree$  to  (0002).  This provides support for the dominant hypothesis for the rationalisation of dwell fatigue crack nucleation in Ti alloys, which derives from the Stroh pile-up model, and elaboration of the underlying dislocation phenomena that result from load shedding and lead to basal faceting.
\end{abstract}
\begin{keyword}
Titanium alloy \sep Dislocations \sep Electron microscopy \sep Fatigue \sep Creep  %% keywords here, in the form: keyword \sep keyword
\end{keyword}

\end{frontmatter}

%% Start line numbering here if you want
%\linenumbers

%% main text	
\section{Introduction}
Sustained holds in load, so-called dwell fatigue, can lead to reductions in the cyclic life of titanium alloys and are therefore of concern to the jet engine industry~\cite{bache2003review, whittaker2011titanium}. The issue first arose as far back as the 1970s and its avoidance almost certainly leads to the over-design of components with a consequent increase in fuel consumption. It is a complex problem as it involves the understanding of plasticity, creep and fracture, and their interaction with microstructure, stress state and stressed volume. Extensive research has been carried out in the past and the metallurgical factors affecting cold dwell fatigue are understood in some detail~\cite{chan1981deformation,luquiau1997cyclic, singh2002low, xiao2002cyclic, singh2007low, li2007comparison, huang2011cyclic, wu2013effect} . 

Fractographic observations have shown that the failure occurs by facet nucleation~\cite{evans1994dwell,bache1997dwell,dunne2007systematic,dunne2007lengthscale, dunne2008mechanisms}, believed to initiate from a strongly misoriented grain pair. These are grain pairs with a particular crystallographic orientation where a `soft' grain well oriented for pri$\langle a \rangle$ slip lies adjacent to a `hard' grain poorly oriented for pri$\langle a \rangle$ slip. This is termed a soft/hard grain pair due to the combined elastic and plastic anisotropy between grains with their $c$-axes near-perpendicular and parallel to the principal applied stress.   A facet forms by crack opening on a crystallographic plane generally found to be on or near the basal planes of the hard grain~\cite{hasija2003deformation,sinha2006observations, bache2010experimental}. Cracking has been observed on facet-like features having orientations 0-$16\degree$ to the basal planes, during both low cycle (fast) and dwell fatigue~\cite{davidson1980titanium, sinha2006crystallography}.     

High local stresses developed by the dislocation pile-ups observed in the soft grain are held to be responsible for this facet nucleation~\cite{hasija2003deformation, dunne2007lengthscale, dunne2008mechanisms}. These stresses are large when the misorientation between grains is large since slip transfer across the boundary is not then possible. During load holds, redistribution of stress from the soft grain to the adjacent hard grain occurs as a function of time, termed load shedding~\cite{hasija2003deformation, dunne2007lengthscale, dunne2008mechanisms}. 

Room temperature creep is thought to be responsible for load shedding (e.g. by Hasija et al.\cite{hasija2003deformation}). Their crystal plasticity modeling predicted a large strain accumulation in the hard grain due to load shedding from the soft grain.  This is a kind of transient creep process in which creep rate decreases continually with time. This creep is not expected to be associated with diffusion mediated mechanisms such as dislocation climb, as the deformation occurs in the vicinity of room temperature. TEM studies show that dislocation glide in the form of planar slip is responsible~\cite{neeraj2001short}. Discrete dislocation plasticity (DDP) studies by Zheng et al. \cite{zheng2017mechanistic} support the notion that significant load shedding and hence dwell sensitivity is expected when the time constant associated with dislocation escape is comparable to the duration of the stress dwell. This process is further expected to be associated with differential strain rate sensitivity of different slip systems~\cite{hasija2003deformation, jun2016nanoindentation, zhang2016intrinsic}.

The 1954 Stroh pile-up model~\cite{stroh1954formation} has been extensively used by the research community to rationalise facet nucleation. It was further modified and applied to understand facet formation in titanium alloys by Bache~\cite{bache1999processing, bache2003review} and Evans and Bache~\cite{evans1994dwell}, as it provides the crack opening stress in a grain due to dislocation pile-ups in a neighbouring grain. The facet is expected to form on a plane having maximum tensile stress; in addition slip on the cracking plane is necessary for cracks to develop~\cite{evans1994dwell, wojcik1988stage}. 

Even though this analytical model of load shedding is well developed and extensively used, being a continuum model it does not provide insight into the associated dislocation structures. It is unknown how  load redistribution from the soft grain nucleates dislocations in the hard grain, the mechanisms responsible for creep deformation and thereby crack nucleation. The nucleation step is believed to be the most important, as titanium alloys are generally notch sensitive. In our recent studies, we observed facet crack nucleation in Ti6242 alloy by near-basal plane splitting due to the large tensile stress developed by a double ended pile-up under low cycle fatigue~\cite{joseph2018dislocation}. TEM studies on the same alloy showed that $\langle a\rangle$ prism pile-up in the soft grain nucleated non-connected $\langle a\rangle$ dislocations in the hard grain under LCF~\cite{joseph2018slip}. DDP studies showed the nucleation of basal slip in the hard grain due to the higher stresses generated by the strong pile-ups in the soft grain under dwell~\cite{zheng2016discrete}. However, such experimental work on dislocations has not been reported under dwell fatigue and the DDP calculations are 2D and are therefore unable to capture events such as dislocation cross-slip and dissociation.  

In this work, we attempt to understand the dislocation mechanisms associated with basal faceting on the hard grain due to load shedding by the soft grain. The dislocation mechanisms associated with three different soft/hard grain pairs under dwell conditions have been investigated and compared with those observed in LCF. Grain pairs with slightly different crystallographic misorientations are considered in order to understand the effect of misorientations between the grains on the dislocation mechanisms. 
	
\section{Experimental Description}
The Ti-6Al-2Sn-4Zr-2Mo-0.1Si (wt.\%) alloy investigated in this work was melted from elemental stock and then processed by rolling in both the $\beta$ and $\alpha+\beta$ phase fields, recrystallized in the $\alpha+\beta$ field at $900\celsius$ for $7\usk\hour$ and air-cooled. The alloy was then aged at $593\celsius$ for 8h and air cooled to promote nanometre-scale Ti$_3$Al precipitation. This processing route resulted in a bimodal microstructure of primary alpha ($\alpha_p$) grains and regions of transformed $\beta$. 
	 
Low cycle fatigue (LCF) and dwell fatigue tests were carried out on cylindrical plain fatigue samples $4.5\usk\milli\meter$ in diameter and $15\usk\milli\meter$ in gauge length using a Mayes servohydraulic machine with an Instron 8800 controller. A trapezoidal waveform with a ramp up/down time of $1\usk\second$, a $1\usk\second$ hold at maximum stress of 95\% (988 MPa) of the yield stress, $1\usk\second$ hold at minimum stress and an $R$ ratio of 0.05 was used for the low cycle fatigue tests. A hold period of $120\usk\second$ was applied at the maximum stress for the dwell tests. Figure 1 shows the LCF and dwell fatigue loading cycles. The strain in the gauge section of the sample was recorded during the test using an Epsilon extensometer with a $10\usk\milli\meter$ gauge length. The tests were carried out until sample failure.

\begin{figure}[t]
\begin{center}\includegraphics[width=90mm]{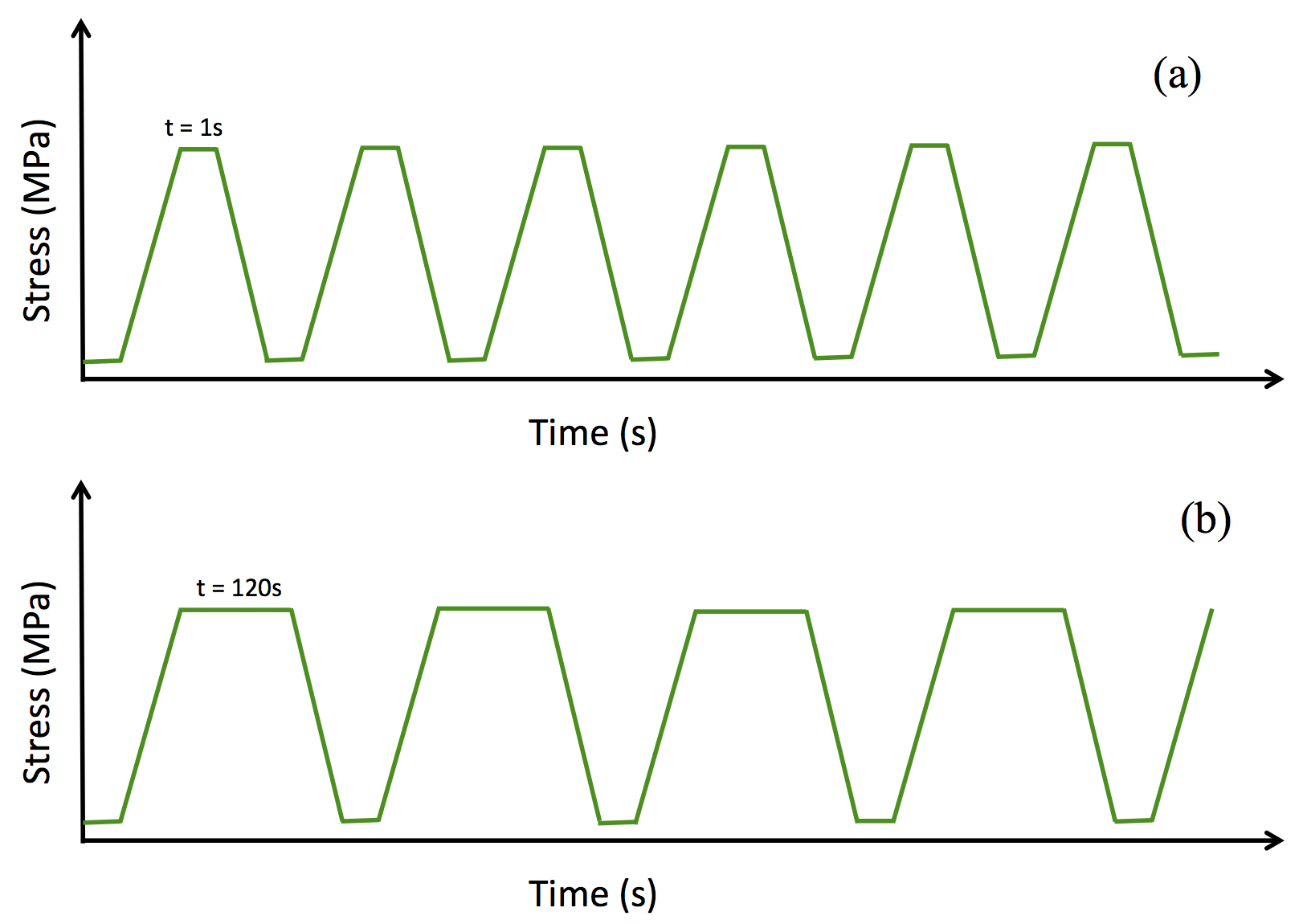}
\end{center}
\caption{The applied load cycles in (a) LCF and (b) dwell fatigue. The maximum stress applied was $95\%$ of the yield stress and the R ratio was 0.05.}
\end{figure}

A Zeiss Sigma300 field emission gun scanning electron microscope (FEG-SEM) in back scattered electron imaging mode was used for initial microstructural analysis. Dislocation analysis was conducted on failed samples using a JEOL JEM-2100F TEM/STEM with an accelerating voltage of $200\usk\kilo\volt$. Discs 0.5 mm thick were cut from the gauge sections of the failed samples, normal to the loading direction, ground to a thickness of $100-150$ $\mu$m using SiC paper and electropolished using $3\%$ perchloric acid, $57\%$ methanol and $40\%$ butan 1-ol in a Struers Tenupol at $-40\celsius$ and 24V.

Transmission Kikuchi Diffraction (TKD) was carried out on TEM foils to identify the grain pair of interest for the dislocation analysis. The data was collected on the same Zeiss Sigma300 SEM used for microscopy with an accelerating voltage of $30\usk\kilo\volt$, working distance of $3\usk\milli\meter$ and the sample in a TKD holder, normal to the electron beam.

\section{Results}

\subsection{Initial microstructure}
Figure 2a shows the bimodal microstructure of the alloy with $\alpha_p$ grains (light phase) in the transformed $\beta$ (dark phase). The volume fraction of $\alpha_p$ was found to be 64\%.  Very thin secondary alpha ($\alpha_s$) platelets were observed in the retained $\beta$, Figure 2b. The typical texture of the alpha regions obtained from EBSD scan showed a strong (0001) texture, Figure 3 (c and d). 

\begin{figure}[t]
\begin{center}\includegraphics[width=90mm]{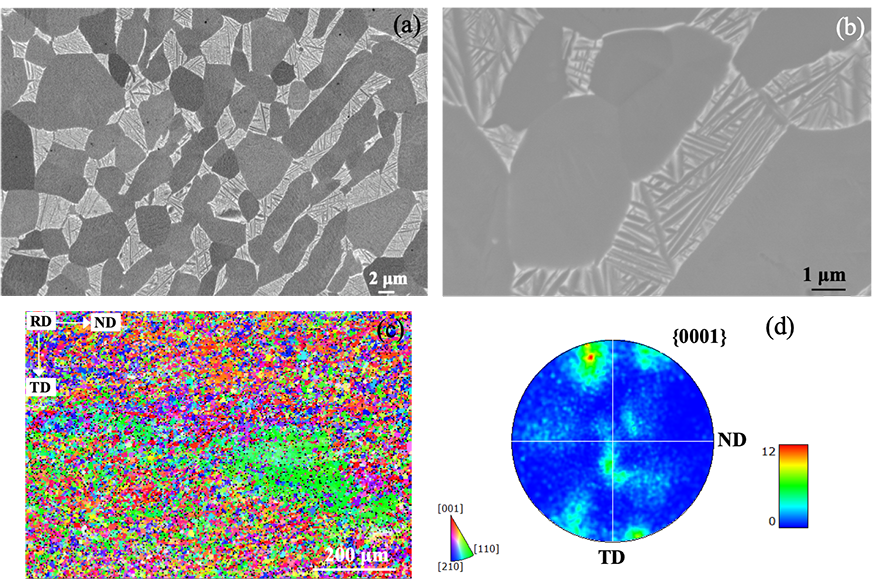}
\end{center}
\caption{Backscattered electron image showing the (a) initial microstructure of Ti6242Si alloy investigated, (b)  Secondary alpha ($\alpha_s$) platelets in retained $\beta$, (c) Large area EBSD scan showing texture in the alloy, (d) (0001) pole figure of $\alpha_p$ phase obtained from the scan in Figure c.}
\end{figure}

\subsection{Mechanical test results}
Tensile testing showed that the alloy possessed a $0.2\%$ yield strength of about 1040 MPa. Accumulation of macroscopic strain was much more pronounced during dwell fatigue than in LCF, Figure 3, with a lifetime to failure of 13359 cycles under LCF and 8655 under dwell conditions. The dwell debit (life ratio) was therefore 1.54, which is relatively low for a near-$\alpha$ Ti alloy.  This is attributed to the relatively small prior-$\beta$ and hence macrozone size associated with small bar processing compared to that associated with large, multi-ton billet.

\begin{figure}[t]
\begin{center}\includegraphics[width=90mm]{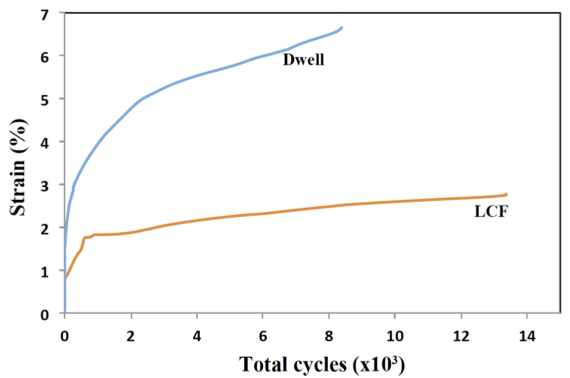}
\end{center}
\caption{Strain accumulation (maximum in each fatigue cycle) during low cycle and dwell fatigue in Ti 6242Si alloy. Higher strain accumulation was observed under dwell conditions. The initial strain of ~0.8\% is a consequence of the nominally elastic straining of the sample.}
\end{figure}

\subsection{Dislocation interactions in the soft/hard grain pairs}
Dislocation analysis was carried out on the TEM foils taken from the gauge section of the failed LCF and dwell samples. Dislocations were analyzed in both bright field (BF) and weak beam dark field (WBDF) imaging mode under two-beam conditions. Each grain in the foils was tilted to at least three different beam directions $B$ and three different $g$ vectors under each beam condition in order to analyze the dislocations. $g.b$ invislibility analysis to identify the Burgers' vector of dislocations will be presented for example grains. Scanning transmission electron microscopy (STEM) was used to capture the dislocation structures in a grain, with the grain tilted to either one of its zone axes, or to the two beam condition.

Grain pairs with particular crystallographic orientations were selected for dislocation analysis, Figure 4. These are inverse pole figure (IPF) maps with respect to the loading direction, obtained by TKD of the TEM foils. Three grain pairs were chosen in each case (LCF and dwell); their $c$-axis orientations to the loading direction are shown in Table 1. The grain pairs were selected to obtain very similar orientations for the comparison of the effect of LCF versus dwell on the dislocation mechanisms occurring. Grain pair 3 is the worst grain pair in each case, having the maximum misorientation of the $c$-axes of the soft and hard grains, which were orientated approximately $85-90\degree$ and $4-8\degree$ to the loading direction, respectively.

\begin{figure}[b!]
\begin{center}\includegraphics[width=90mm]{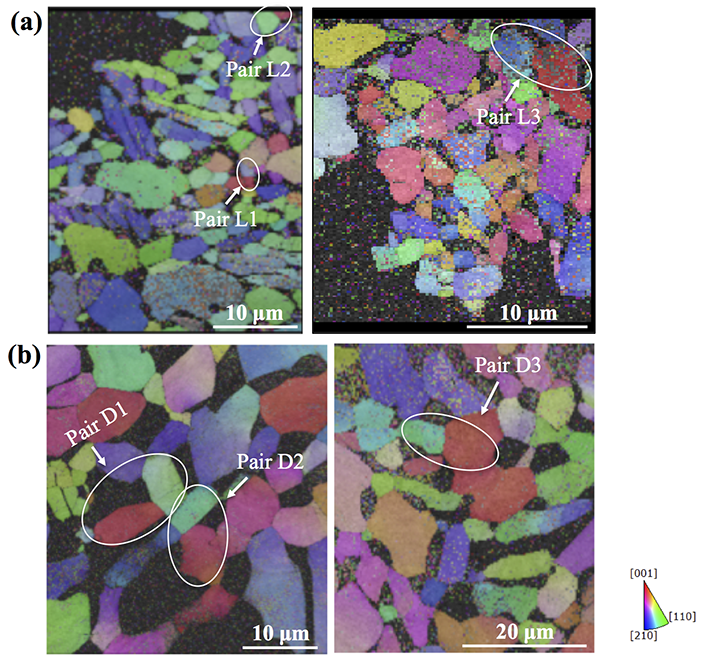}
\end{center}
\caption{Transmission Kikuchi Diffraction images (loading axis out of page, IPF coloured with respect to the out-of-plane direction) showing the grain pairs investigated from the gauge section of the samples loaded in (a) LCF [pairs L1-L3] and (b) dwell [pairs D1-D3].}
\end{figure}

\begin{table}\centering
\begin{tabular}{cllll}\hline
Grain pair & LCF & & Dwell & \\
 & Soft & Hard & Soft & Hard \\\hline
1 & 70.0 & 13.4 & 73.8 & 8.5\\
2 & 80.2 & 20.0 & 85.3 & 20.6\\
3 & 84.7 & 3.6 & 89.5 & 7.9 \\\hline
\end{tabular}
\caption{$c$-axis orientation (in degrees) of the grain pairs analysed, with respect to the loading direction.}
\end{table}

\subsection{Low cycle fatigue}
\begin{figure*}[t!]
\begin{center}\includegraphics[width=150mm]{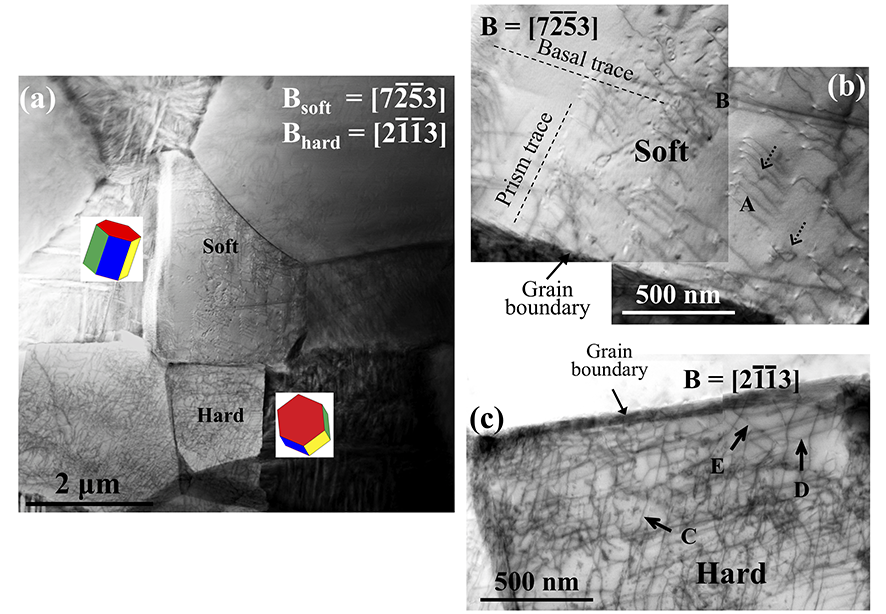}
\end{center}
\caption{Bright field STEM composite micrograph showing (a) overall dislocation structures across the soft/hard grain pair L1 under LCF; high magnification images showing (b) $\langle a \rangle$ type prism pile-ups in the soft grain and (c) arrays of long straight dislocation segments on basal plane in the hard grain. The beam directions and grain orientations to the loading axis are inset; the loading axis was normal to the plane of the foil.}
\end{figure*}
We first consider the dislocation structures observed in the three grain pairs from the LCF sample. Figure 5a shows the overall dislocation structures observed in grain pair L1. This is a BF-STEM composite micrograph with the soft grain tilted to $B = [7\bar{2}\bar{5}3 ]$ and the hard grain then tilted to $B=[2\bar{1}\bar{1}3 ]$. BF-STEM imaging permits observation of all the dislocations simultaneously, except for those with line directions parallel to the beam, and is relatively insensitive to bend contours and other imaging artefacts. The high magnification image in Figure 5b shows the dislocation pile-ups in the soft grain more clearly. The major slip systems, A and B, are highlighted. $g.b$ invisibility analysis shows that the piled-up dislocations in slip system A are of $(a/3)[2\bar{1}\bar{1}0]$ type, gliding on the prism plane. These dislocations showed cross-slip and dislocation loop formation (similar to the letter $\alpha$ in appearance), indicated by arrows. The dislocations in slip system B are found to be of $(a/3)[\bar{1}2\bar{1}0]$ basal type. These dislocations are very few in number and some of these types of dislocations at the bottom of the grain are interacting with the boundary on the right hand side and not interacting with the hard grain of interest, so these will not be discussed further. Figure 5c shows arrays of intersecting or overlapping dislocations observed in the hard grain. These are different $\langle a\rangle$ type dislocations gliding on basal planes, indicated by slip systems C, D and E. The Burgers vectors of these dislocations are listed in Table 2.  

\begin{figure*}[t!]
\begin{center}\includegraphics[width=150mm]{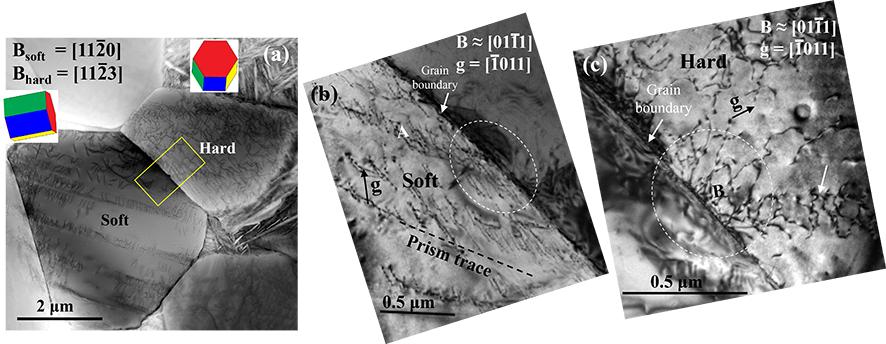}
\end{center}
\caption{(a) Bright field STEM composite micrograph showing overall dislocation structures across soft/hard grain pair L2 in the LCF sample, (b) BF-TEM image showing pile-ups of dislocations in the soft grain and (c) BF-TEM image showing nucleation of numerous dislocations from the boundary in the hard grain. The beam conditions are inset and the loading direction is normal to the foil. Micrographs b and c were captured under two-beam conditions.}
\end{figure*}
The dislocation interactions between the soft and hard grains of pair L2 are shown in Figure 6a. This is again a BF-STEM composite micrograph, when the soft grain is tilted to $B = [11\bar{2}0]$ and the hard grain is then tilted to $B = [11\bar{2}3]$.  It is worth noting that the dislocations in the pile-ups of the soft grain in Figure 6a are grouped on both ends of the pile-up - e.g. that the pile-ups are double- rather than single-ended.  The dislocations near the boundary can be seen more clearly under a two-beam condition in Figures 6(b, c). These are high magnification images taken from the region highlighted by a box in Figure 6a.The same regions in the micrographs of Figures 6(b, c) are highlighted by dotted circles. It can be seen that there are numerous dislocations (slip system B) emerging from the boundary in the hard grain, Figure 6c where the dislocation pile-ups (slip system A) in the soft grain impinge, Figure 6b. These are $\langle a\rangle$ type prism pile-ups in the soft grain and $\langle a\rangle$ type basal dislocations in the hard grain. A different pile-up can be seen in the hard grain, indicated by an arrow in Figure 6c; the curvature of these dislocations suggests that they are hitting the boundary. This pile-up is not discussed further as we are interested in the dislocations nucleating from the boundary of a hard grain. The dislocation types and habit planes are listed in Table 2.

\begin{figure*}[t!]
\begin{center}\includegraphics[width=150mm]{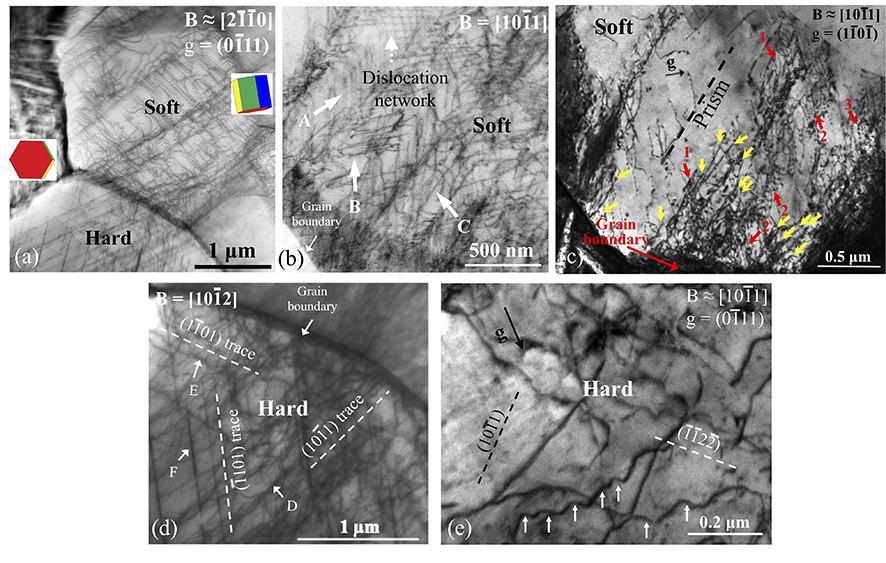}
\end{center}
\caption{BF-STEM (a) micrograph showing overall dislocation structures in the most misoriented grain pair L3 under LCF, with the soft grain in the two-beam condition $B\approx[2\bar{1}\bar{1}0]$ and g = $(0\bar{1}11)$, (b) Dislocation structures in the soft grain when $B = [10\bar{1}1]$,  (c) BF-TEM image showing the pile-ups (slip system A) and long dislocation lines (slip system C) in the soft grain under two beam condition, (d) Slip systems observed in the hard grain when $B=[10\bar{1}2]$ and (e) BF-TEM image showing cross-slip of $\langle c+a \rangle$ dislocations in the hard grain in slip system D under a two-beam condition. The beam conditions are inset. The loading axis was normal to the plane of the foil.}
\end{figure*}
The BF-STEM image in Figure 7a shows the dislocation interactions in the most-misoriented grain pair L3. This micrograph was captured when the soft grain was tilted to a two beam condition, $B\approx [2\bar{1}\bar{1}0]$ and g = $(0\bar{1}11)$ showing the dislocation interactions between the grains. The piled-up dislocations (slip systems A and B) and the long dislocation lines (slip system C) observed in the soft grain are shown in Figure 7b, with the grain tilted to the beam condition $B=[10\bar{1}1]$. The dislocation pile-up in slip system A and the long dislocation lines in slip system C are shown more clearly under two beam condition $B\approx [10\bar{1}{1}]$ and g = $(1\bar{1}0\bar{1})$, Figure 7c. The red arrows show the pile-ups and yellow arrows show the long dislocation lines.  All the dislocations in the soft grain were found (by $g.b$ invisibility and trace analysis) to be of $\langle a\rangle$ type, gliding on prism planes. The piled-up dislocations are found to be of screw character impinging the boundary, whereas the long dislocations are found to be of edge character, and all these dislocations are observed to touch the boundary, meaning that they nucleated from the boundary. In addition, dislocation networks were also observed in the soft grain (dotted arrow in Figure 7b). 

\begin{table*}[t!]\centering
%\begin{adjustbox}{angle=90}
\begin{small}
%\hspace{-3cm}
\begin{tabular}{c | lll | lll}\hline
&& Soft& & & Hard & \\\hline
\rotatebox{90}{Grain pair}&\rotatebox{90}{Slip system}& \rotatebox{90}{Burgers vector} & \rotatebox{90}{Slip plane} &\rotatebox{90}{Slip system}& \rotatebox{90}{Burgers Vector} & \rotatebox{90}{Slip plane}\\\hline
L1 & Pile-up (A) & $(1/3)[2\bar{1}\bar{1}0]$ & $(01\bar{1}0)$ & C& $(1/3)[\bar{2}110]$ & (0001) \\
& Long dislocation (B)& $(1/3)[\bar{1}2\bar{1}0]$ & $(0001)$ & D& $(1/3)[\bar{1}2\bar{1}0]$ & (0001)   \\
& &  &  & E & $(1/3)[11\bar{2}0]$ & (0001) \\\hline
L2 & Pile-up (A) & $(1/3)[\bar{2}110]$ & $(01\bar{1}0)$ & B & $(1/3)[11\bar{2}0]$ & (0001)  \\\hline
 L3& Pile-up(A) & $(1/3)[\bar{2}110]$ & $(01\bar{1}0)$ & D & $(1/3)[11\bar{2}\bar{3}]$ & Cross-slip between $(\bar{1}\bar{1}2\bar{2})$ and $(10\bar{1}1)$  \\
 & Pile-up (B) & $(1/3)[11\bar{2}0]$ & $(1\bar{1}00)$ & E & $(1/3)[\bar{1}2\bar{1}3]$ & Cross-slip between  $(1\bar{1}01)$ and $(1\bar{2}12)$ \\
 & Nucleated (C) & $(1/3)[\bar{1}2\bar{1}0]$ & $(10\bar{1}0)$ & F & $(1/3)[\bar{2}11\bar{3}]$ &  $(\bar{1}101)$ \\\hline
\end{tabular}
\end{small}
%\end{adjustbox}
\caption{Slip systems observed in the grain pairs of the LCF sample.}
\end{table*}
Figure 7d shows slip systems D, E and F observed in the hard grain when the grain is tilted to $B=[10\bar{1}2]$. It could be seen that the slip systems D and F are emerging from the boundary. $g.b$ invisibility analysis shows that these are $\langle c+a\rangle$ type dislocations gliding on first order pyramidal planes. We have observed cross slip of dislocations in slip systems D and E, with their glide planes confirmed from tilting experiments. The high magnification image under two beam condition in Figure 7(e) show cross-slip of dislocations in one of the slip systems D, indicated by arrows. These dislocations are found to cross-slip between the first and second order pyramidal planes. No cross-slip was observed in slip system F. 

This grain pair L3 showed the nucleation of dislocations from the boundary in the soft grain in addition to dislocation nucleation in the hard grain, in contrast to the other grain pairs L1 and L2. The Burgers vector and the slip planes of the various slip systems observed in the investigated grain pairs of the LCF samples are summarised in Table 2. 

\subsection{Dwell fatigue}
\begin{figure*}[t]
\begin{center}\includegraphics[width=150mm]{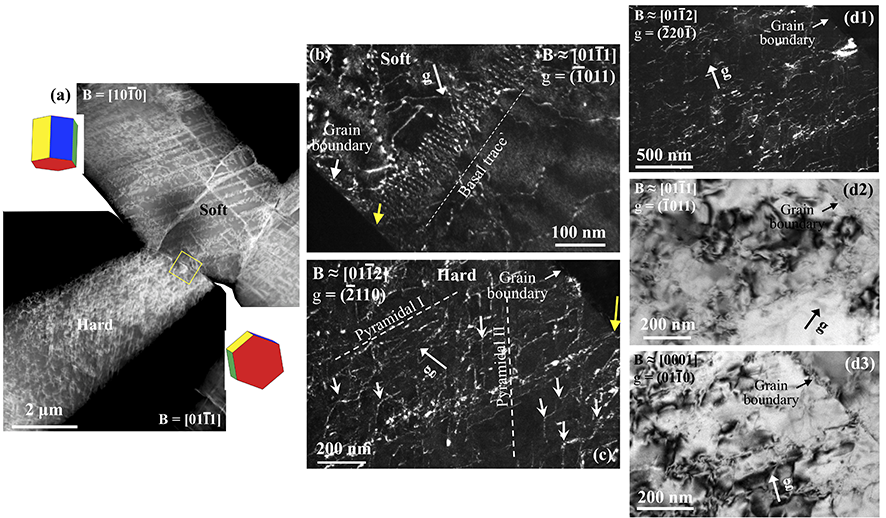}
\end{center}
\caption{(a) DF-STEM composite micrograph showing overall dislocation structures in soft/hard grain pair D1 in dwell fatigue, when the soft grain was tilted to $B = [10\bar{1}0]$ and the hard grain was tilted to $B = [01\bar{1}1]$, (b) WBDF image showing pile up of $\langle a\rangle$ dislocations impinging on the boundary in the soft grain and (c) WBDF image showing nucleation of $\langle c+a\rangle$ dislocations from the boundary in the hard grain under two-beam conditions. Cross-slip of $\langle c+a\rangle$ dislocations are indicated by arrows. (d1-d3) $g.b$ invisibility analysis for Burgers vector identification of dislocations in the hard grain under two beam conditions. The beam conditions are inset and the loading axis was normal to the plane of the foil.}
\end{figure*}
We now turn to consider the dislocation observations in the dwell sample. The DF-STEM image in Figure 8a shows the dislocation interactions in grain pair D1. This is a composite micrograph of dark field STEM images when the soft grain is tilted to the beam direction $B = [10\bar{1}0]$ and the hard grain is tilted to the beam direction $B = [01\bar{1}1]$. High magnification weak beam dark field $(g-3g)$ images are shown in Figures 8(b--c). These images were taken from the region highlighted by a box in Figure 8a. The same region in the micrographs are indicated by yellow colored arrows. It can be seen that the dislocations are emerging in the hard grain from the point where the pile-up in the soft grain hits the boundary, shown by yellow arrows in Figs. 8(b--c). The g.b invisibility analysis show that these are $\langle a\rangle$ basal dislocations in the soft grain and $\langle c+a\rangle$ pyramidal dislocations in the hard grain. Thus an $\langle a\rangle$ basal pile-up in the soft grain resulted in the nucleation of $\langle c+a\rangle$ dislocations in the hard grain.  There were other dislocations in the soft grain in addition to this pile-up, which are not analysed here. The $\langle c+a\rangle$ dislocations in the hard grain were found to cross-slip between the first and second order pyramidal planes, indicated by arrows in Figure 8c. The dislocation density in the hard grain was found to be higher than in LCF, at least qualitatively from the observation of the foils in different scattering orientations. g.b invisibility analysis for the identification of Burgers vector of the dislocations in the hard grain (in Figure 8(c)) is shown in Figure 8(d1-d3). The dislocations are visible under g=$(\bar{2}20\bar{1})$ and invisible under g=$(\bar{1}011)$ and g=$(01\bar{1}0)$, resulting in the inference that they possessed a Burgers vector of $b = (a/3)[2\bar{1}\bar{1}3]$. The Burgers vector and the habit plane of these dislocations are listed in Table 3.
\begin{table*}[t!]\centering
%\begin{adjustbox}{angle=90}
\begin{small}
%\hspace{-3cm}
\begin{tabular}{c | lll | lll}\hline
&& Soft& & & Hard & \\\hline
\rotatebox{90}{Grain pair}&\rotatebox{90}{Slip system}& \rotatebox{90}{Burgers vector} & \rotatebox{90}{Slip plane} &\rotatebox{90}{Slip system}& \rotatebox{90}{Burgers Vector} & \rotatebox{90}{Slip plane}\\\hline
L1 & Pile-up (A) & $(1/3)[2\bar{1}\bar{1}0]$ & $(01\bar{1}0)$ & C& $(1/3)[\bar{2}110]$ & (0001) \\
& Long dislocation (B)& $(1/3)[\bar{1}2\bar{1}0]$ & $(0001)$ & D& $(1/3)[\bar{1}2\bar{1}0]$ & (0001)   \\
& &  &  & E & $(1/3)[11\bar{2}0]$ & (0001) \\\hline
L2 & Pile-up (A) & $(1/3)[\bar{2}110]$ & $(01\bar{1}0)$ & B & $(1/3)[11\bar{2}0]$ & (0001)  \\\hline
 L3& Pile-up(A) & $(1/3)[\bar{2}110]$ & $(01\bar{1}0)$ & D & $(1/3)[11\bar{2}\bar{3}]$ & Cross-slip between $(\bar{1}\bar{1}2\bar{2})$ and $(10\bar{1}1)$  \\
 & Pile-up (B) & $(1/3)[11\bar{2}0]$ & $(1\bar{1}00)$ & E & $(1/3)[\bar{1}2\bar{1}3]$ & Cross-slip between  $(1\bar{1}01)$ and $(1\bar{2}12)$ \\
 & Nucleated (C) & $(1/3)[\bar{1}2\bar{1}0]$ & $(10\bar{1}0)$ & F & $(1/3)[\bar{2}11\bar{3}]$ &  $(\bar{1}101)$ \\\hline
\end{tabular}
\end{small}
%\end{adjustbox}
\caption{Slip systems observed in the grain pairs of the LCF sample.}
\end{table*}

Figure 9 shows a DF-STEM image of grain pair D2 with the soft grain tilted to $B = [10\bar{1}1]$. The high magnification images in Figures 9(b--c) are taken from the region highlighted by a box in Figure 9(a). Figure 9(b) shows the long dislocation lines observed in the soft grain. There are pile-ups in this grain which are not impinging the hard grain of interest and not discussed. In contrast to all the LCF cases (e.g. Figure 6a) where dislocation pile-ups were observed, in dwell (Figure 9a) long, straight dislocation segments in the soft grain impinge directly on the hard grain boundary. 
It could be seen that some dislocation loops (highlighted by yellow colored dotted lines) started from the same region of the boundary where these long dislocation lines impinge, Figure 9(c). This implies that the impinging dislocations caused the nucleation of dislocation loops in the hard grain. The long dislocation lines in the soft grain were found to be $\langle a\rangle$ type basal dislocations with edge character. The dislocation loops in the hard grain were found to be of $\langle c+a\rangle$ pyramidal type. The dislocation density in this hard grain is not as high as in the previous grain pairs. These slip systems are listed in Table 3.   

\begin{figure*}[t!]
\begin{center}\includegraphics[width=150mm]{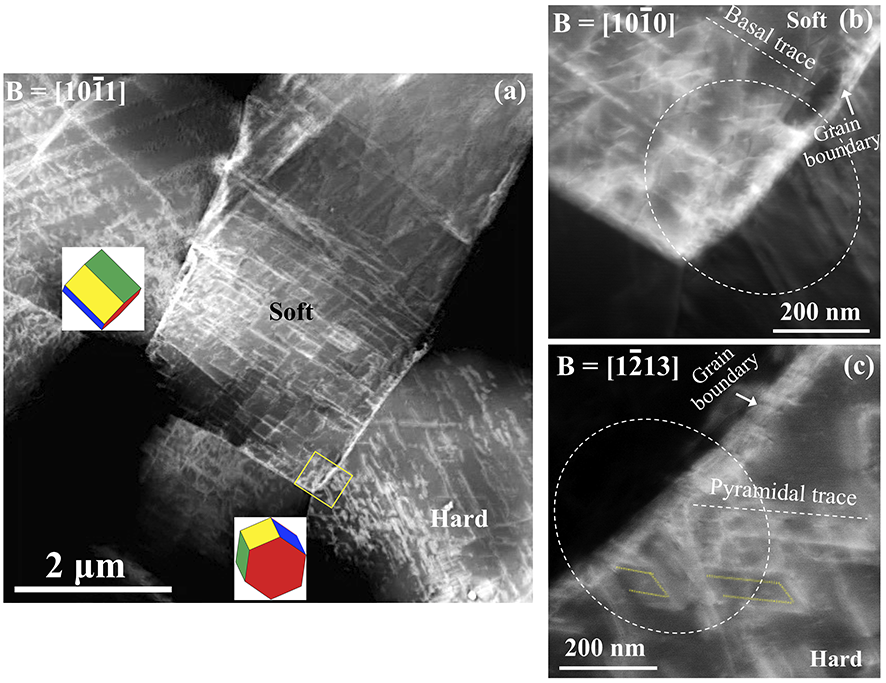}
\end{center}
\caption{DF-STEM micrograph showing (a) overall dislocation structures in soft/hard grain pair D2 when the soft grain is tilted to $B = [10\bar{1}1]$ under dwell fatigue, (b) long dislocation segments impinging the boundary in the soft grain when $B = [10\bar{1}0]$ and (c) nucleation of $\langle c+a\rangle$ pyramidal loops from the boundary in the hard grain when $B = [1\bar{2}13]$. The loading axis was normal to the plane of the foil.}
\end{figure*}

The overall dislocation structures observed near the boundary of the most misoriented grain pair D3 are shown in Figures 10(a--b). A large number of slip could be seen in both soft and hard grains. The various slip systems in the soft grain are shown in Figures 10(c--d). The pile-ups in slip systems A and B were found to impinge upon the boundary (Figure 10c). A very few long dislocation lines  were found in the soft grain in addition to pile-ups (not analysed). There are some long dislocation lines (slip system C) starting from the boundary in the soft grain (Figure 10d).  The various slip systems D, E and F in the hard grain (Figure 10e) were found to emerge from the boundary. The boxes in Figure 10a highlight the regions from which the images in Figure 10(c-e) are taken. The pile-up dislocations in slip systems A and B were found to be of $(a/3)[11\bar{2}0]$ and $(a/3)[\bar{2}110]$ type with screw character, gliding on prism planes $(1\bar{1}00)$ and $(01\bar{1}0)$ respectively.  Slip system C starting from the boundary was of $(a/3)[\bar{1}2\bar{1}0]$ prism-type with edge character. 

\begin{figure*}[t!]
\begin{center}\includegraphics[width=150mm]{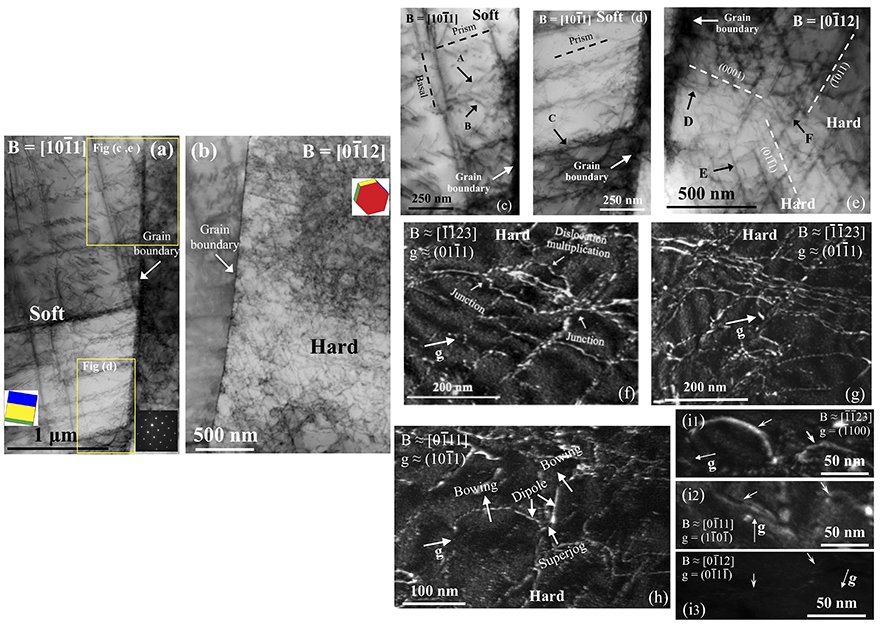}
\end{center}
\caption{BF-STEM micrograph showing the overall dislocation structures near the boundary in the (a) soft grain, and (b) the hard grain in the most misoriented grain pair D3 of the dwell sample. High magnification BF-STEM image showing (c) $\langle a\rangle$ prism pile-ups in the soft grain impinging the boundary, (d) nucleation of $\langle a\rangle$ prism dislocations with edge character from the boundary in the soft grain, (e) nucleation of numerous dislocations from the boundary in the hard grain, WBDF images from the hard grain showing (f) basal dislocation multiplication by junction formation, (g) higher dislocation density on the basal plane  (h) dislocation source activation by superjog formation in slip system F and (i1-i3) g.b invisibility analysis for Burgers vector identification of dislocations of (h) under two beam conditions. The beam conditions are inset and the loading axis was normal to the plane of the foil.}
\end{figure*}

The high density of dislocations observed in the hard grain were found to be basal (slip system D) and pyramidal (slip system E and F) dislocations. $g.b$ invisibility analysis showed that the basal dislocations are $(a/3)[\bar{1}2\bar{1}0]$ type and the pyramidal dislocations are of $(a/3)[11\bar{2}3]$ and $(a/3)[\bar{1}2\bar{1}\bar{3}]$ types.  The formation of dislocation junctions and subsequent dislocation multiplication was observed in the basal slip system, Figure 10(f), which leads to a high density of basal dislocations in the hard grain, Figure 10(g). Formation of superjogs and the bowing of adjacent dislocation segments was observed in both pyramidal slip systems F and G. This observation is shown for one of the slip systems F in Figure 10(h). $g.b$ invisibility analysis for the identification of the Burgers vector of the dislocations of Figure 10(h) is shown in Figures 10(i1-i3). These are weak beam dark field images taken under two beam conditions. The bowing dislocation segments are indicated by arrows. The dislocations are visible under g=$(\bar{1}100)$ and invisible under g=$(1\bar{1}0\bar{1})$ and g=$(0\bar{1}1\bar{1})$ resulting in a Burgers vector of $b = (a/3)[\bar{1}2\bar{1}\bar{3}]$. The slip systems observed in the grain pairs of the dwell samples are summarised in Table 3. In general, the dislocation density in the hard grains was found to be higher in the dwell sample than in those of the LCF sample, except grain pair D2.

\subsection{Schmid factor values}
The activation of a slip system in a particular grain depends on the shear stress acting on the grain; calculated Schmid factors (neglecting grain interaction stresses and considering only the remote load) therefore are often helpful in rationalising the observed slip systems. The highest values for the different slip systems under LCF and dwell fatigue are shown in Table 4.

\begin{table*}\centering\begin{small}\addtolength{\tabcolsep}{-2pt}  
\begin{tabular}{c | ccccc | ccccc}
& & &Soft& & & &Hard & &  &   \\\hline
\rotatebox{90}{Grain Pair}& \rotatebox{90}{$\langle a\rangle$ Basal }& \rotatebox{90}{$\langle a\rangle$ Prism} & \rotatebox{90}{$\langle a\rangle$ Py I} & \rotatebox{90}{$\langle c+a\rangle$ Py I} & \rotatebox{90}{$\langle c+a\rangle$ Py II} & \rotatebox{90}{$\langle a\rangle$Basal} & \rotatebox{90}{$\langle a\rangle$Prism} & \rotatebox{90}{$\langle a\rangle$ Py I} & \rotatebox{90}{$\langle c+a\rangle$ Py I} & \rotatebox{90}{$\langle c+a\rangle$ Py II} \\\hline
LCF& & & & & & & & &  \\
L1 & \textcolor{red}{0.304} &	\textcolor{red}{0.437}	&0.499&	0.491&0.435&\textcolor{red}{0.205}&	0.025&	0.116&	0.483&	0.495\\
L2 & 0.166&	\textcolor{red}{0.475}&	0.467&	0.481&	0.485&	\textcolor{red}{0.290}&	0.054&	0.178&	0.484	&0.481\\
L3 & 0.087&	\textcolor{red}{0.491}	&0.464&	0.471	&0.432&	0.063	&0.002&	0.030&	\textcolor{red}{0.435}&	\textcolor{red}{0.474}\\\hline
Dwell&&&&&&&&&&\\
D1 & \textcolor{red}{0.263}&	0.457&	0.485&	0.490&	0.479&	0.143&	0.007&	0.072&	\textcolor{red}{0.468}&	\textcolor{red}{0.493}\\
D2 & 0.080&	\textcolor{red}{0.493}&	0.459&	0.466&	0.461&	0.290&	0.055&	0.184&	\textcolor{red}{0.480}&	0.477\\
D3 & 0.009&	\textcolor{red}{0.499}&	0.441&	0.440&	0.431	&\textcolor{red}{0.134}&0.009&0.067&\textcolor{red}{0.465}& \textcolor{red}{0.492}\\\hline
\end{tabular}\\
\caption{Schmid factor values on grain pairs under LCF and dwell fatigue. The operative slip systems observed by TEM are highlighted in \textcolor{red}{red}.}
\addtolength{\tabcolsep}{2pt}  
\end{small}
\end{table*}

In general, slip systems with high Schmid factor values were found to activate as expected. However, a slip system with lower Schmid factor value was observed in the grain pair D3 of dwell sample: basal slip in the hard grain. This slip system must therefore have activated as a consequence of slip or strain transfer from the neighbouring grain. 

\subsection{Pile-up stress}

Here we develop a mathematical description of the stress concentration generated at the hard/soft grain boundary due to dislocation pile-up in the soft grain, to evaluate the hypothesis that this may be responsible for crack nucleation in the hard grain. The Stroh pile-up model provides a quantitative expression for the normal stress on an inclined plane due to a dislocation pile-up~\cite{stroh1954formation}. Stroh's original model considered a remotely applied pure shear stress parallel to the pile-up plane and a pile-up composed of edge dislocations. However, in our case the applied stress is tensile and the dislocations in the pile-up are of screw character. Hence Stroh's results cannot be used directly to estimate the stress in the present situation, and so here we extend that model to the present case.

As the screw dislocations generate out-of plane shear stresses the problem is considered in 3D. Let us now consider the case for worst grain pair where the soft grain has its $(10\bar{1}0)$ prism plane perpendicular to the loading direction. This configuration leads to dislocation pile-ups on the other two prism planes, $(01\bar{1}0)$ and $(1\bar{1}00)$, which would make an angle of 30$\degree$ to the loading axis, shown schematically in Figure 11a for one configuration. A tensile stress $\sigma_0$ is applied in the z-direction and the cracking plane in the hard grain makes an angle $\theta$ to the pile-up plane. 

\begin{figure}[t]
\begin{center}\includegraphics[width=90mm]{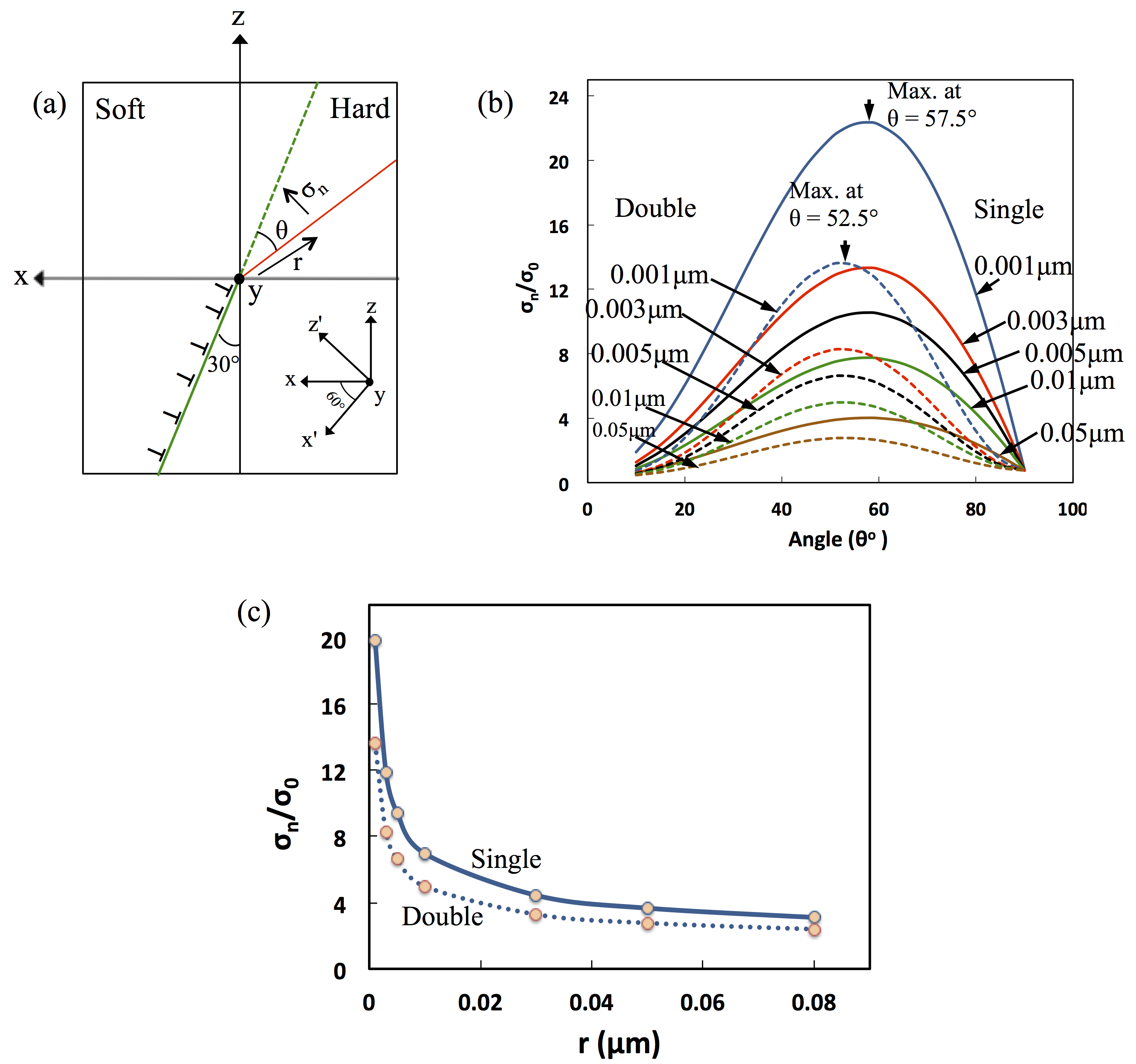}
\end{center}
\caption{(a) Schematic showing the pile-up model, (b) variation of normalized stress with angle $\theta$ at a fixed distance $r$ and (c) variation of normalized stress with distance $r$ from the front of the pile-up at a fixed angle $\theta$ }
\end{figure}

The length of the pile-up~\cite{eshelby1951xli} is given by $L = \frac{Gbn}{\pi\tau_{0}}$, where $G$ is the shear modulus, $b$ is the Burgers vector, $n$ is the number of screw dislocations in the pile-up and $\tau_{0}$ is the applied shear stress.

The normal stress $\sigma_n$ acting on the plane in the hard grain, which makes an angle $\theta$ with the pile-up plane, is then found to be
\begin{equation}
 \begin{aligned}
\frac{\sigma_{n}}{\sigma_0} = \frac{{1}}{{4}} \left(\sqrt{3}\sin\theta+\cos\theta\right)^2+\left(\frac{3L}{4r}\right)^\frac{1}{2}\sin\frac{\theta}{2}\sin2\theta
\end{aligned}
\end{equation}

where $r$ is the distance from the front of the pile-up. This result is derived in the Appendix.

The length of the pile-up becomes $L=\frac{D}{2}$ when the dislocation pile-up is single ended and the source is at the center of the grain, where D is the grain length. Then the normal stress developed by the single ended pile-up is given by
\begin{equation}
 \begin{aligned}
\frac{\sigma_{n_{s}}}{\sigma_0} = \frac{{1}}{{4}} \left(\sqrt{3}\sin\theta+\cos\theta\right)^2+\left(\frac{3D}{8r}\right)^\frac{1}{2}\sin\frac{\theta}{2}\sin2\theta
\end{aligned}
\end{equation}

In contrast, when the pile-up becomes double-ended the length of the pile-up becomes $L=\frac{D}{4}$ and the source is at the center of the grain. The factor of 4 instead of 2 arises because the back stress on the source arises from dislocations piled-up on both sides of the source. Therefore the normal stress developed by a double ended pile-up is given by
\begin{equation}
 \begin{aligned}
\frac{\sigma_{n_{d}}}{\sigma_0} = \frac1{4}\left(\sqrt{3}\sin\theta+\cos\theta\right)^2+\left(\frac{3D}{16r}\right)^\frac{1}{2}\sin\frac{\theta}{2}\sin2\theta
\end{aligned}
\end{equation}

The variation of normal stress with $\theta$ is shown in Figure 11(b). The normal stress found to be at a maxima when $\theta=57.5\degree$ for a single ended pile-up and $\theta=52.5\degree$ for a double ended pile-up. Figure 11(c) shows that the normal stress is found to be inversely proportional to $\sqrt{r}$ for a fixed hard grain inclination angle $\theta$. These are the maximum normal stress values (at $\theta = 57.5\degree$ for a single ended pile-up and $\theta=52.5\degree$ for a double ended pile-up as predicted from Figure 11b)  and a reduction of $\approx 34\%$ of stress for a double ended pile-up is expected compared to a single ended pile-up immediately adjacent to the boundary, at $r=0.001\usk\mu\meter$. A grain size (D) value of $5\usk\mu\meter$ is taken for these calculations which is the average grain size of the alloy studied.   

Hence, for the single ended pile-ups observed in the dwell samples, the maximum stress plane would be $57.5\degree$ to the pile-up plane. The pile-up plane itself is at an angle of $30\degree$ to the loading direction, to maximise its Schmid factor for prism $\langle a\rangle$  slip. Hence the maximum stress plane in the hard grain would be $\approx 87.5\degree$ to the loading direction, $2.5\degree$ from $\{0002\}$. The basal plane in the neighboring hard grain is almost $90\degree$ to the loading direction which means that the maximum stress plane would be near-basal for dwell fatigue. Similarly, the maximum stress plane would be $\approx 82.5\degree$ to the loading direction for LCF since double ended pile-up was observed. Hence the maximum stress plane would be $\approx 7.5\degree$ to the basal plane in LCF. However, it should be cautioned that this analysis does not take into account lattice rotations due to plasticity, so this difference may not provide a unique experimental measure to distinguish LCF from dwell in failure analysis. More sophisticated analytical approaches to similar problems exist which might provide further insight, such as those of Le and coworkers \cite{berdichevsky2007dislocation, kochmann2008dislocation, le2008analytical, kochmann2009plastic}, or indeed computational modelling approaches such as DDP \cite{dunne2007lengthscale,zheng2016discrete}.

\section{Discussion}

In general, the dislocation density in the grain pairs was found to be higher in dwell fatigue than in LCF.  Numerous dislocation pile-ups were observed in all the soft grains analysed in both low cycle fatigue and dwell fatigue due to the favorable orientation of these grains for slip. These kinds of pile-ups are expected in this alloy due to the slip planarity that results from $\alpha_{2}$ precipitation~\cite{ joseph2018dislocation, joseph2018slip, savage2001deformation}. The pile-ups observed in the soft grain were $\langle a\rangle$ type prism slip in most cases. The pile-up density was observed to be higher in the worst grain pair with the dislocations piling-up on two prismatic planes, Figures 7b and 10c. 

Cross-slip events and loop formation were observed in these dislocations, Figure 5b, indicating that these dislocations were generated by multiple cross-slip events as we have previously observed~\cite{joseph2018slip}. This kind of planar slip produces only minimal strain hardening. Further, such cross-slip events can result in large amounts of strain, which is also consistent with the low strain hardening rate~\cite{neeraj2000phenomenological}. This type of extreme planar slip was previously observed in $\alpha$ Ti-6Al in creep~\cite{neeraj2000phenomenological}. Hence, it is anticipated that the soft grains creep during the initial load cycles, with the creep rate decreasing with time as these dislocation pile-ups impinge the grain boundary. This kind of creep is expected when the time constant associated with dislocation escape is comparable to the duration of stress hold \cite{zheng2017mechanistic}.   

The pile-ups in the soft grain were found to be double ended in LCF, Figures 6a and 7c. We have previously observed this kind of double ended pile-up in this near-$\alpha$ Ti alloy (in a different microstructural condition) under LCF ~\cite{joseph2018dislocation, joseph2018slip}. A double-ended pile-up can result from multiple cross-slip events~\cite{joseph2018slip, kacher2016situ} and/or incomplete reversibility of dislocation motion due to a slightly higher friction stress in the reverse loading than in the forward loading ~\cite{joseph2018dislocation, tanaka1981dislocation}.  Importantly, double ended pile-ups were not observed under dwell loading. 
\begin{figure*}[t!]
\begin{center}\includegraphics[width=150mm]{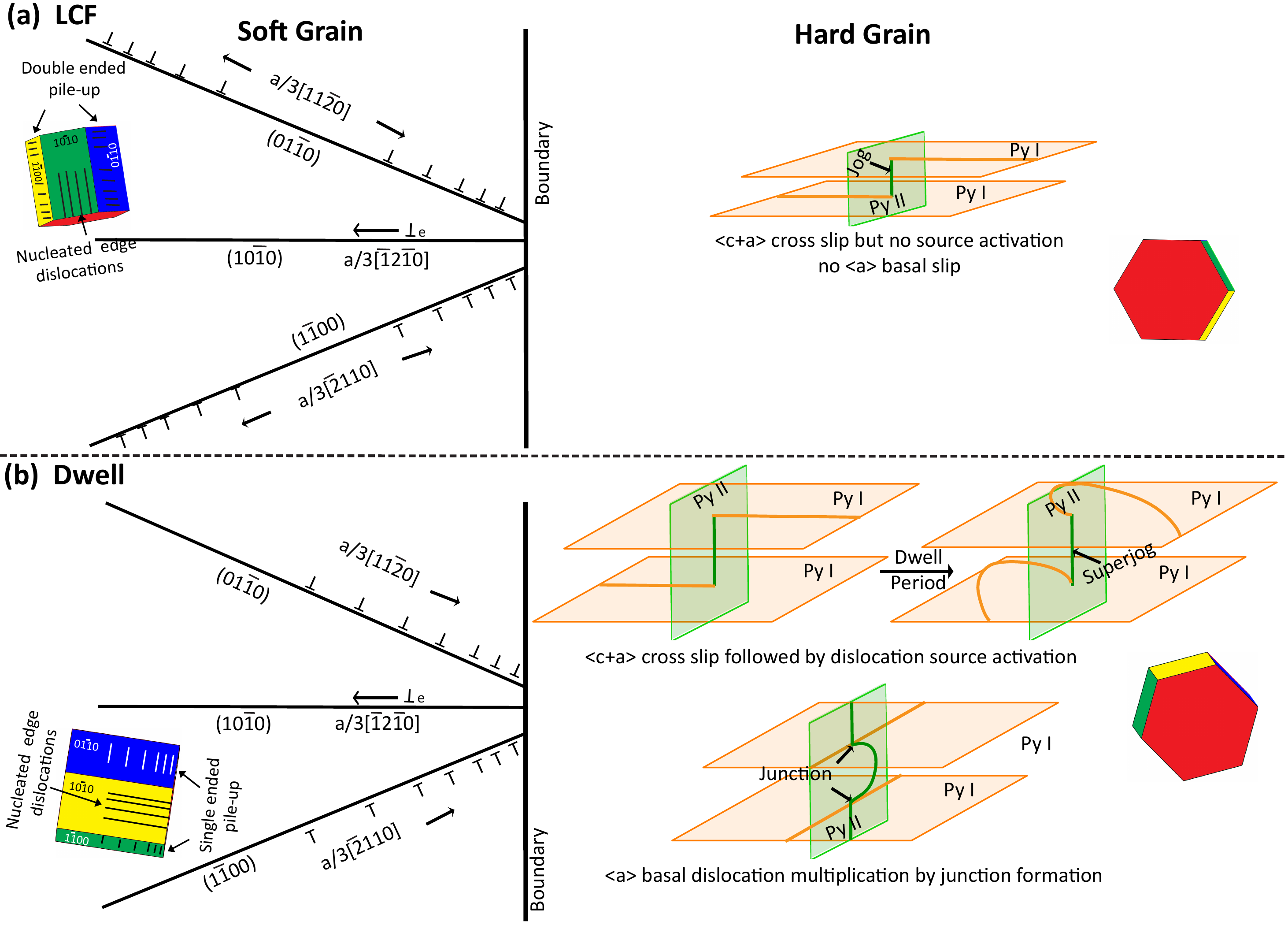}
\end{center}
\caption{Schematic showing the observed dislocation mechanisms in the worst grain pair in near-$\alpha$ Ti6242Si, with 64\% $\alpha_p$ under (a) LCF, where the pile up stress is lower in the double-ended case, and (b) dwell fatigue with a single-ended pile up, shedding more load onto the hard grain than in the case of LCF.}
\end{figure*}

An important observation was made for more highly misorientated hard/soft grain pair under dwell, for example, when the misorientation is $>80\degree$. The soft grain in this pair showed pile-up on two prism planes, $(01\bar{1}0)$ and $(1\bar{1}00)$, Figure 10(c). This is possible since the nearly perpendicular orientation of the $[0002]$ axis means that these two prism planes possessed almost equal Schmid factor values (dual slip). The slip transmission of these pile-ups to the neighbouring hard grain is highly unfavourable due to the large misorientation between the grains. This must lead to a high stress concentration near the pile-up front. In addition, we have observed slip producing edge dislocations on the third prism plane $(10\bar{1}0)$, Figure 10(d), which is hard to deform. This slip is expected to activate as a consequence the high local stresses developed by the other two prism pile-ups. The nucleation of these dislocations from the boundary confirms this. This third pile-up will then result in an additional pile-up stress concentration in the neighbouring hard grain. These pile-ups near the boundary then activated a large density of dislocations from the boundary in the hard grain, Figure 10(e). This observation was also made in LCF for grain pair L3, Figure 7. 

Disconnected dislocations, long dislocation arrays and dislocation loops were observed in the hard grain. A higher dislocation density was also observed in the hard grains than in the soft, even though the hard grains were unfavorably orientated for slip. The hard grain dislocation density was found to be significantly greater in the dwell case than in LCF.  This suggests that dislocation generation in the hard grain is favored by the internal stresses developed during deformation and the increased amount of time available for load/strain shedding, as a consequence of the stress concentration developed by the soft grain pile-ups. Our pile-up stress calculations show that this stress would be higher (by $\approx 34\%$) for the single ended pile-ups observed under dwell fatigue than in for the double ended pile-ups observed in LCF.

Both $\langle a\rangle$ basal and $\langle c+a\rangle$ pyramidal slip were observed to nucleate from the boundary in the hard grains.  The $\langle c+a\rangle$ dislocation showed cross-slip between the first and second order pyramidal planes whereas cross-slip was not observed in $\langle a\rangle$ basal dislocations. This cross-slip could be seen in LCF pair L3 (Figure 7) and dwell pairs D1 and D3 (Figures 8  and 10). This cross-slip was most obvious when the mis-orientation between the grains was high, and in dwell fatigue. Such cross-slip events will allow for large amounts of strain in the hard grain.

$\langle c+a\rangle$ cross-slip in the hard grain of worst grain pair resulted in superjog formation under dwell fatigue, Figure 10. Superjogs form when the stress is high enough to string multiple jogs along the dislocation lines, having a step height of 5 to $30b$. Dislocation segments on either side of the jog then bow out when the shear stress is increased. As the step height of these superjogs is greater than $\approx20$ nm, the distance between the dislocation segments is large enough to prevent mutual interaction. In this case, the dislocations behave as single-ended dislocation sources~\cite{george1988dieter, oh2009situ}. In contrast, cross-slip was observed in the LCF case but there was no activation of a dislocation source. Therefore it is inferred that the source was activated under dwell because the critical stress for source activation could be achieved as a result of time-dependent load shedding from the soft grain during the load holds~\cite{zheng2016discrete}. 

It is worth noting that a high density of basal dislocations was observed in the hard grain under dwell fatigue, Figure 10(g), which was not observed in LCF. These basal dislocations are difficult to activate in the hard grain purely from the remote loading, due to the low Schmid factor values, Table 4, and are therefore a consequence of activation by load shedding from dislocation pile-ups in the soft grain during load hold.  Dislocation interactions are nearly inevitable when there are numerous dislocations~\cite{bulatov2006dislocation}; basal dislocations in the hard grain were found to form junctions by interacting with other dislocations, Figure 10(f). The dislocation segments between these junctions then multiply under dwell, leading to the high basal dislocation density observed. 

The schematic in Figure 12 depicts the overall dislocation mechanisms associated with facet nucleation for a highly misoriented grain pair, for both LCF and dwell fatigue loading. There are dislocation pile-ups on two prismatic planes, $(01\bar{1}0)$ and $(1\bar{1}00)$, in the soft grain. These pile-ups were found to be single ended for dwell fatigue and double ended for LCF. The local stress developed by these pile-ups generated slip on the third prism plane, $(10\bar{1}0)$. Thus, there would be a large number of prism pile-ups and slip transmission is not possible due to high misorientation between the grains which resulted in large pile-up stress. The stress concentration developed by these pile-ups would be higher in the dwell scenario due to the observed single ended pile-up. This stress concentration then results in the nucleation of numerous dislocation sources in the hard grain, by load shedding. Superjog formation by cross-slip of $\langle c+a \rangle$ dislocations leads to source activation and subsequent dislocation multiplication in the dwell case. The junction formation mechanism leads to basal dislocation multiplication. These superjog and junction  dislocation multiplication mechanisms were not observed in LCF. 

Hence substantial strain accumulation occurs in the hard grain in dwell by $\langle a \rangle$ basal and $\langle c+a \rangle$ pyramidal slip. In addition, a greater tensile stress can be developed by the single ended pile-up in the soft grain acting on the hard grain. Our analytical calculations predicts that the maximum stress plane would be near-basal ($\approx 2.5\degree$ to $(0002)$) under dwell fatigue. Thus, the high tensile stress and shear on the basal plane might plausibly result in quasi-cleavage of the basal plane and thereby basal faceting under dwell fatigue. The cracking plane would be very similar, $\approx 7.5\degree$ to $(0002)$, under LCF.

\section{Conclusions}

The dislocation interactions between soft/hard grain pairs in near-$\alpha$ Ti6242Si were investigated under low cycle and dwell fatigue in order to understand the effect of load holds on the dislocation mechanisms leading to crack nucleation. The following conclusions are drawn:

1. A higher density of dislocations was observed in the grain pairs during dwell fatigue than in LCF, increasing with mis-orientation between the grains. The density was found to be higher in the hard grain than in the soft grain, which is suggested to be a consequence of the internal stress developed by prism $\langle a\rangle$ pile-ups in the soft grain. 

2. The pile-up stress was derived for a set of screw dislocations in the soft grain, and the normal stress developed on the hard grain as a result of these pile-ups was evaluated, giving an expression for the stress available for facet nucleation. This stress was found to be higher in dwell fatigue than in LCF due to the single-ended pile-ups observed under dwell conditions.

3.  In the most highly misoriented pairs, a high density of prism $\langle a\rangle$ pile-ups was observed by dual slip due to the favourable orientation of the soft grain for slip. The local stress developed by these pile-ups activated slip on the hard prism plane. The pile-up stress concentration in this case would be higher due to a greater number of prism pile-ups and the unfavourabilty of slip transmission between the grains, which resulted in nucleation of numerous dislocations in the hard grain.

4. Both $\langle a\rangle$ basal and $\langle c+a\rangle$ pyramidal dislocations were observed in the hard grain under dwell fatigue. The $\langle c+a\rangle$ pyramidal dislocations were found to multiply by superjog formation. The $\langle a\rangle$ basal dislocations multiplied by junction formation and resulted in high density of basal dislocations. $\langle a\rangle$ basal dislocations were not observed and $\langle c+a\rangle$ dislocations were found to cross-slip, but these did not subsequently multiply under LCF.

5. The higher pile-up stress and basal dislocation multiplication in the hard grain are expected to be associated, following Stroh [25] and Tanaka and Mura [35], with basal faceting and early crack nucleation during dwell fatigue.

%\newpage 
\section*{Acknowledgements}
%\vspace{-1em}
\noindent\small The authors wish to acknowledge the contribution and useful discussions with Prof D Rugg at Rolls-Royce plc, Prof FPE Dunne and Dr TB Britton at Imperial, and Prof AJ Wilkinson in Oxford, with whom we are funded under the Hexmat EPSRC programme grant EP/K034332/1.

%\section*{References}
%\bibliographystyle{unsrt}
%\small\setlength{\itemsep}{0cm} \bibliography{dwell}

%\newpage
\section*{A. Appendix}

Here, the normal stress in an adjacent hard grain due to an applied normal stress on a screw dislocation pile-up in a neighbouring soft grain is derived.   The applied stress in the $xyz$ coordinate system is 

\begin{equation}
\sigma=
\begin{pmatrix}
0 & 0 & 0  \\
0 & 0 & 0 \\
0 & 0 & \sigma_0
\end{pmatrix}
\end{equation}

In the $x'y'z'$ system (when the rotation is $60\degree$ counter-clockwise about $y$), the applied stress tensor becomes

\begin{equation}
\sigma'=\frac{\sigma_0}{4}
\begin{pmatrix}
3 & 0 & -\sqrt{3}  \\
0 & 0 & 0 \\
-\sqrt{3}  & 0 & 1
\end{pmatrix}
\end{equation}

The resolved shear stress on the pile-up plane is therefore $\tau_{0}=-\sqrt{3}\sigma_0/4$.

Consider a pile-up of $n$ screw dislocations in the soft grain. The leading dislocation is locked by the grain boundary and the remaining $(n-1)$ dislocations are free to move on the slip plane under the applied shear stress $\tau_0$. The equilibrium positions of the dislocations can be obtained from the zeros of the derivative of the $n^{th}$ Laguerre polynomial $L_{n}'(Z)$~\cite{eshelby1951xli}. Then the stress due to these $(n-1)$ dislocations is given by

\begin{equation}
\sigma_{y'z'} + i\sigma_{x'z'} =\frac{Gb}{2\pi} \frac{F'(Z)}{F(Z)}
\end{equation}

where G depends on the direction of Burgers vector and slip plane, F(Z) is the derivative of the $n^{th}$ Laguerre polynomial and Z=X+iY.   
For large $n$, the asymptotic expansion of $L_n'(Z)$ is given by~\cite{stroh1954formation}
\begin{equation}
L_{n}'(Z) =\frac{{1}}{{2}}K\pi^{-\frac{{1}}{{2}}}e^\frac{{Z}}{{2}}(-Z)^{-\frac{{3}}{{4}}}n^{-\frac{{1}}{{4}}}e^{2(-nZ)^{1/2}[1+O(n^{-1/2})]}
\end{equation}  
where K is a constant equal to exp($2\tau_{0}-\frac{Gb}{2\pi}$). $(-Z)^\frac{1}{2}$ is taken to be real and positive when Z is real and negative and the length of the dislocation pile-up is taken as L=4n, after~\cite{stroh1954formation}.

From equations (6) and (7), the stress due to the $(n-1)$ dislocations are
\begin{equation}
\begin{aligned}
&\sigma_{x'z'} =\frac{{Gb}}{{2\pi}} \left[-\frac{{3}}{{4r}}\sin\theta+\left(\frac{n}{r}\right)^\frac{1}{2}\sin\frac{{\theta}}{{2}}\right]\\
&\sigma_{y'z'} =\frac{Gb}{2\pi} \left[\frac{{3}}
{{4r}}\cos\theta-\left(\frac{n}{r}\right)^\frac{1}{2}\cos\frac{{\theta}}{{2}}+\frac{{1}}{{2}}\right]
\end{aligned}
\end{equation}
In addition, there are stresses due to the locked dislocation and the applied stress which are given by
\begin{equation}
\begin{aligned}
&\sigma_{x'z'} =\frac{-Gb\sin\theta}{2\pi r} \\
&\sigma_{y'z'} =\frac{Gb\cos\theta}{2\pi r} 
\end{aligned}
\end{equation}
From Equations (8) and (9), the total stress due to the dislocation pile-up is
\begin{equation}
 \begin{aligned}
&\sigma_{x'z'} =\frac{{Gb}}{{2\pi}} \left[-\frac{{7}}{{4r}}\sin\theta+\left(\frac{n}{r}\right)^\frac{1}{2}\sin\frac{{\theta}}{{2}}\right] \\
&\sigma_{y'z'} =\frac{Gb}{2\pi} \left[\frac{{7}}{{4r}}\cos\theta-\left(\frac{n}{r}\right)^\frac{1}{2}\cos\frac{{\theta}}{{2}}+\frac{{1}}{{2}}\right]
\end{aligned}
\end{equation}

The $\frac{1}{r}$ term can be neglected in the above equation as $r>\frac{1}{n}$ and $\frac{1}{r}<(\frac{n}{r})^\frac{1}{2}$. Therefore, Equation (10) becomes

\begin{equation}
\begin{aligned}
\sigma_{x'z'} &=\frac{Gb}{2\pi} \left[\quad\left(\frac{n}{r}\right)^\frac{1}{2}\sin\frac{\theta}{2}\right] \\
\sigma_{y'z'} &=\frac{Gb}{2\pi} \left[-\left(\frac{n}{r}\right)^\frac{1}{2}\cos\frac{\theta}{2}+\frac{1}{2}\right]
\end{aligned}
\end{equation}

The total stress is the sum of the pile-up stress and the applied stress which is given by 

\begin{equation}
\sigma'=
\begin{pmatrix}
   \frac{3}{4}\sigma_0  & 0 & \sigma_{x'z'}-\frac{\sqrt[]{3}}{4}\sigma_0   \\
   0 & 0 & \sigma_{y'z'} \\
  \sigma_{z'x'}-\frac{\sqrt[]{3}}{4}\sigma_0  & \sigma_{z'y'} & \frac{\sigma_0}{4} 
\end{pmatrix}
\end{equation}

The normal stress $\sigma_{zz}$ acting in the xyz coordinate system at $(r, \theta)$ is then 
\begin{equation}
 \begin{aligned}
\sigma_{n} = \frac{\sigma_0}{2}\left[\frac{{1}}{{2}} \left(\sqrt{3}\sin\theta+\cos\theta\right)^2+\left(\frac{3L}{r}\right)^\frac{1}{2}\sin\frac{\theta}{2}\sin2\theta\right]
\end{aligned}
\end{equation}
where the first term arises from the applied stress and the second from the pile-up.

\end{document}